# Multi-walled carbon nanotube films for the measurement of the alcoholic concentration

*C. Giordano[1,2], G. Filatrella[1], Maria Sarno[3,4] and A. Di Bartolomeo[2,4,*]*

[1]Department of Sciences and Technologies, University of Sannio, Via Port'Arsa 11, I-82100 Benevento, Italy
[2]Physics Department "E. R. Caianiello", University of Salerno, Via Giovanni Paolo II, 132, I-84084, Fisciano, Salerno, Italy
[3]Industrial Engineering Department, University of Salerno, Via Giovanni Paolo II, 132, I-84084, Fisciano, Salerno, Italy
[4]Interdepartmental Centre NanoMates, University of Salerno, Via Giovanni Paolo II, 132, I-84084, Fisciano, Salerno, Italy
*E-mail: adibartolomeo@unisa.it; Tel.: +39-089-969189

We show that a multi-walled carbon nanotube film can be used as the sensing element of a low-cost sensor for the alcoholic concentration in liquid solutions. To this purpose, we investigate the electrical resistance of the film as a function of the isopropanol concentration in a water solution. The analysis reveals a growing resistance with increasing isopropanol concentration and a fast response. The sensing element is re-usable as the initial resistance value is restored once the solution has evaporated. The electrical resistance increases linearly when the multi-walled carbon nanotube film is exposed to common beverages with increasing alcoholic content.
This work paves the way for the development of low-cost, miniaturized MWCNT-based sensors for quality monitoring and control of alcoholic beverages and general liquid solutions.

**1. Introduction:** The availability of nanostructured materials has generated huge opportunities for the realization of new highly sensitive, low-cost, portable and small-size sensors, which operate at very low power consumption. Carbon nanotubes (CNTs), since they were discovered by Iijima in 1991 [1], have been drawing great research interest because of their one-dimensional morphology and unique chemical, electrical and mechanical properties [2-5]. Thus, the preparation, properties and applications of carbon nanotubes have been the subject of an intense activity. Carbon nanotubes are graphene sheets rolled up to form hollow cylinders with single-walls (SWCNTs) or multi-walls (MWCNTs). SWCNTs have diameters of few nanometers, and length of at least of 1–100 μm [6]. MWCNTs consist of multiple layers of graphene wrapped up around the same central axis to form tubes with diameters up to hundreds nanometers and lengths that can reach several centimeters. The structure of CNT inherently provides unique electrical, physical, and chemical properties. Mechanically, due to the C-C σ-bonds, CNTs are among the strongest fibers that are known to date. CNTs also possess a high thermal stability, both in vacuum and in air. In terms of electrical properties, SWCNTs can be either metallic or semiconducting, depending upon the tube diameter and chirality (the direction in which the graphene sheet is rolled to form the tube) [7], while, for the large diameter and different wall tube chiralities, MWCNTs have a metallic behavior.

The extremely high surface-to-volume ratio makes CNTs ideal for sensing purposes, as the interaction of CNTs with the analyte is maximized [8,9]. Hence, CNTs based sensors have been long sought by the biomedical, automotive, manufacturing, food and fishing industry, as well as for application in agriculture and environmental control [10]. In particular, it has been demonstrated that CNTs are sensitive to gases, which, adsorbed on their surface, easily change their conductivity, capacitance, dielectric constant, etc. The voltage response of the CNTs as a function of time has been used to detect H2 and CO2 at various concentrations, while supplying a constant current to the system [11]. The sensitivity of the response to gas molecules has been related to the extent of electron transfer and the binding energy [12]. SWCNT films have been used for alcohol detection through a change of mass detected by a frequency shift of the SWCNT film deposited onto a quartz crystal microbalance [13]. Alcohol vapor sensors consisting of SWCNT field effect transistors with a significant changes in the FET drain current by exposure to various kinds of alcoholic vapors were reported in ref. [14]. Similarly, ethanol vapor detection was demonstrated using metal–CNT hybrid materials synthesized by infiltrating SWNTs with transition metals such as Ti, Mn, Fe, Co, Ni, Pd or Pt [15]. Alcohol vapor detection was also achieved using MWCNT films, whose resistance was found to increase with the flow rate of the alcohol vapor carriered by an inert gas [16]. Sin et al. [17] demonstrated that the functionalization with carboxyl groups grafted along the sidewall and ends of MWCNTs can increase the variation of the resistance to alcohol vapors. It was proposed that the polar -COOH groups enhance the response to the alcohol vapors, as the dipole-dipole interactions with -COOH increases the absorption efficiency of the volatile polar organic molecules [18].

In 2009, it was reported that the electrical resistance versus temperature of a freestanding MWCNT film, known as buckypaper, has a monotonic behavior, and it was proposed that the buckypaper could be used as a small-size temperature sensor with long term stability, fast response, and high sensitivity on a wide range of temperatures [19].

In this study, differently from previous works, we demonstrate that an easy-to-fabricate MWCNT film can be directly exposed to a liquid solution and exploited as a chemical sensor to monitor the alcoholic concentration. Indeed, the electrical resistance of the buckypaper increases monotonically and reversibly with the alcoholic concentration of the liquid solution. Liquid drops, with different alcohol contents, are easily adsorbed on, or desorbed from, the MWCNT film, and the element responds with a variation of the resistance that is proportional to the alcoholic concentration, enabling the discrimination of liquid solutions based on their alcoholic concentration. Furthermore, we confirm that the MWCNT film is sensitive to temperature (temperature sensor), and has a positive temperature coefficient of resistance ($dR/dT > 0$) typical of metals.

**2. Device fabrication and measurements:** MWCNTs were synthesized by an ethylene catalytic chemical vapor deposition (CCVD) on Co/Fe-Al$_2$O$_3$ catalyst, following a procedure described elsewhere [19,20]. After that, 0.5 g of MWCNTs were suspended in 100 g of water and sonicated for 15 min (see Fig.1) [21]. The solution was then vacuum filtered onto a membrane support. After drying, films of different thickness and densities were removed from the support. This procedure allows some control of the thickness and the density of the buckypaper [20].

The obtained films are stable, freestanding, and bendable; they can be easily manipulated or contacted without breaking.

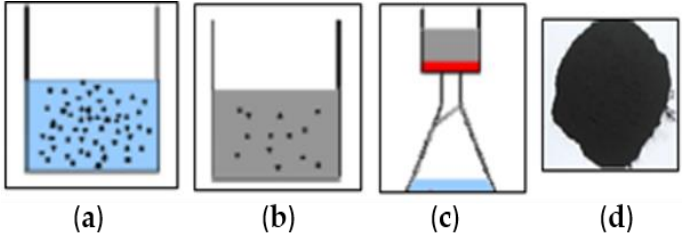

Fig. 1. *Production of a film of MWCNTs (buckypaper). (a) MWCNTs are added to an aqueous solution; (b) A MWCNT dispersion is obtained after sonication; (c) The MWCNT dispersion is filtered under vacuum; (d) Finally a free-standing MWCNTs film is obtained.*

Figures 2 (a) and (b) show an optical and a scanning electron microscope (SEM) image of the buckypaper. The nanotubes are multi-walled with external and the internal diameter in the range from 10 to 30 nm, and from 5 to 10 nm, respectively, as obtained from the transmission electron microscopy characterization reported in a previous work [22].

The buckypaper sample used for the resistance measurement was held by silver paint on a glass support, as shown in Fig. 2 (c) where the electric measurement setup is sketched. It consists of a small rectangle of about 3 cm length, 0.5 cm width and 400 μm diameter. Copper wires (diameter 0.2 mm) were connected to the sample surface by silver paint.

Resistance measurements on the sample kept inside a glass box were performed by a Keithley 4200-SCS using a two- or four-probe configuration, at room temperature and pressure. We forced a current ($I_{force}$) and measured a voltage ($V_{meas.}$), thereby obtaining $R = \frac{V_{meas.}}{I_{force}}$. To avoid Joule heating of the sample, we kept $I_{force}$ below 1 mA (corresponding to less than 3μW power dissipation).

**3. Results and discussion:** We start with a short discussion on the electrical properties and the temperature response of the buckypaper. After that, we focus on the effects of alcohol on the electrical resistance, from which we demonstrate that the buckypaper can be conveniently exploited for the measurement of the alcoholic concentration of an aqueous solution.

*3.1. Electric characterization*

Fig. 3 (a) shows the typical current-voltage (I-V) characteristic of the buckypaper in two- and four-probe configuration at room temperature and pressure. The MWCNT film is highly conductive and the linear I-V behavior confirms its ohmic nature. The electrical resistance, obtained from the slope of the I-V characteristics in Fig. 3 (a), determined with the two probe method is 4.32 Ω, and becomes 2.30 Ω using the four probe method. Hence, we estimate that the contribution of wires and contacts to the total resistance of the sample is about 2 Ω. To focus only on the properties of the buckypaper and to avoid the effect of the contacts, in the following, we systematically use the four-probe configuration that eliminates the contact and the wire resistance.

Fig. 3 (b) shows the electrical resistance as a function of time and evidences that the resistance of the buckypaper is stable over time, since it remains almost constant at 2.26 Ω for 16 h. The slight trend up, starting after 7 hours, corresponds to day/night increase of the temperature of the laboratory. The inset of Fig. 3 (b) displays the increasing electrical resistance for growing temperature resulting from a hot plate placed at different distances (denoted as position 1 and 2) from the sample; the final decay corresponds to the removal of the hot plate. It is clear that the warming up of the sample yields an increase in the resistance.

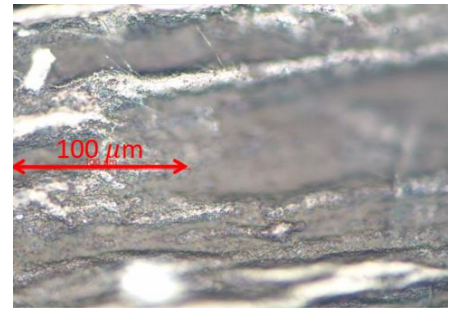

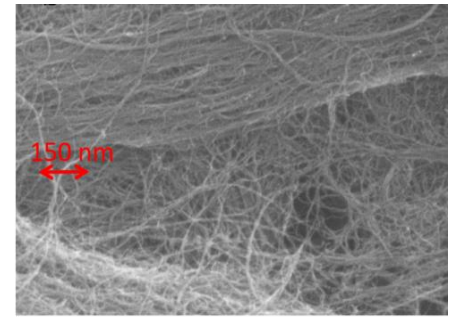

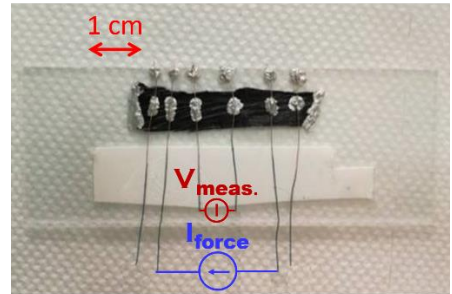

Fig. 2. *SEM and TEM images of the sample. (a) Optical image of the sample; (b) SEM image of the buckypaper showing a network of nanotubes bundles; (c) The sample used in the experiments contacted by six copper wires. Four wires were used simultaneously during the alcoholic test concentration experiment for the four-probe resistance measurement, in which a current was forced between the intermediate wires while a voltage was measured between the innermost ones.*

The resistance dependence on temperature is a further indication of the metallic behavior of the buckypaper under study, which is a direct consequence of the high density of MWCNTs in it [19].The fast response and high sensitivity to temperature is a remarkable advantage of the buckypaper and can be exploited to realize a small-size thermistor with a temperature coefficient of resistance around $7 \cdot 10^{-4} K^{-1}$ as we reported in a previous publication [19].

*3.2. Alcoholic sensor*

In the following, we consider the sample as an alcohol sensor by studying its response to isopropanol and then to isopropanol/water solutions. Fig. 4 shows the time evolution of the resistance when an isopropanol drop (0.05 ml) was poured on the portion of the buckypaper between the two contacts used for the voltage sensing in the four-probe measurement configuration. The small volume assures that the drop is confined away from the two leads and there is no reaction of the analyte with the silver paint contacts. The figure shows that the electrical resistance is suddenly increased by the presence of isopropanol, and that it slowly decreases while the isopropanol evaporates. We point out that the buckypaper and the isopropanol liquid drop are at the same temperature (room temperature), therefore the sudden increase of resistance cannot be ascribed to a drop-induced temperature change. The decrease follows an exponential law, as shown by the fitting curve. The electrical resistance returns to the baseline value after few hours. It is important to notice that the sensors recovery time, i.e. the time needed for the resistance to return to the baseline, is dominated by the slow evaporation of the alcohol. Such time can be shortened by heating the sample by an increased forcing current, which speeds up the evaporation process.

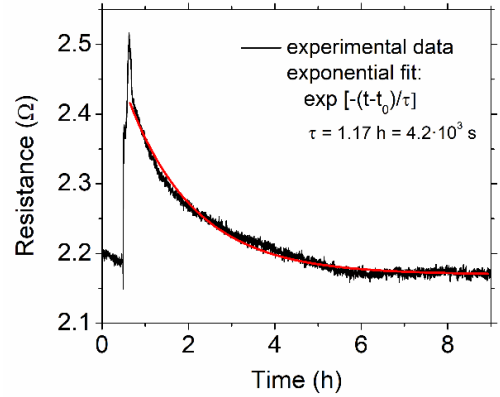

Fig. 4. *Resistance of the buckypaper vs time coated with a 0.05 ml isopropanol drop. The sample is highly sensitive to the isopropanol, which causes a sudden increase of its resistance. The resistance slowly returns to the baseline while the isopropanol evaporates.*

Fig. 5 (a) shows the electrical resistance vs time obtained by pouring an increasing amount of isopropanol on the buckypaper, determined by the number of drops.

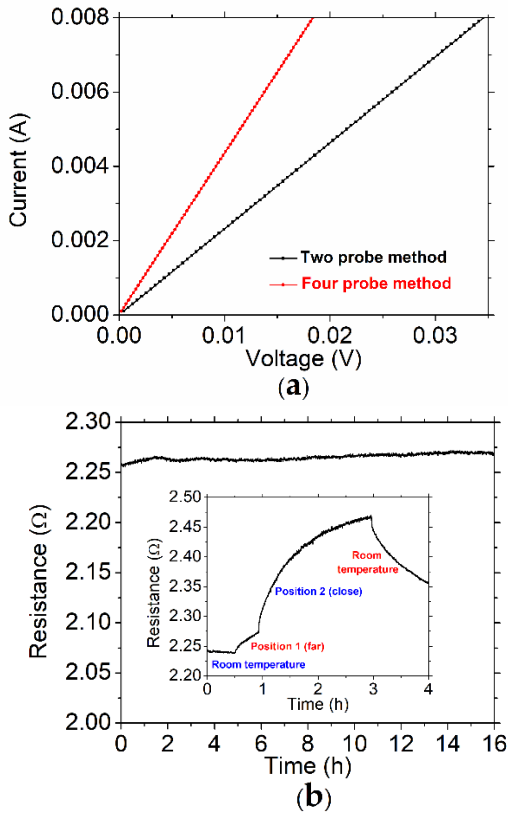

Fig. 3. *(a) Current-Voltage (I-V) characteristics of the buckypaper measured in two (upper red line) and four-probe (lower black line) configurations.(b) Four-probe electrical resistance of the buckypaper as a function of the time. The resistance stays constant for several hours; the fluctuations are due to minor room-temperature variations. The inset shows the response of the sample to temperature changes induced by a hot plat, placed at different distances from the buckypaper. The sample electrical resistance is highly sensitive to temperature.*

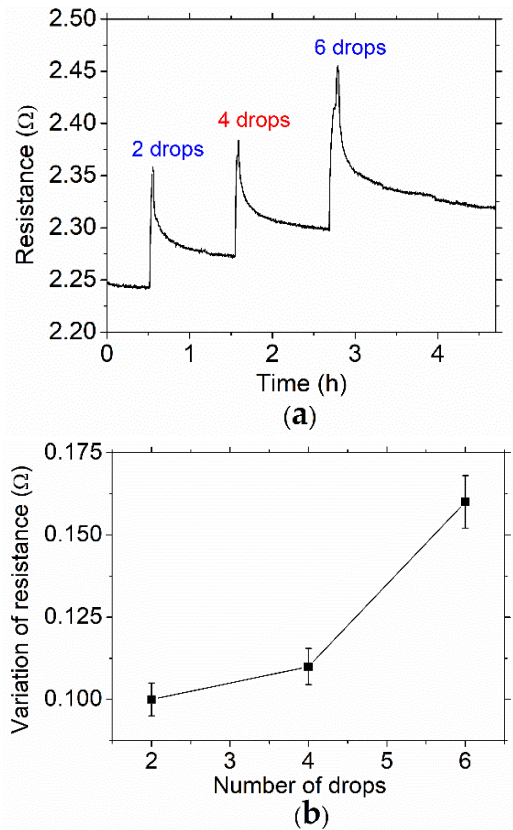

Fig. 5. *Sensor response to increasing number of isopropanol drops. (a) Electrical resistance vs time, showing that adding isopropanol drops causes a sudden increase of the resistance; (b) The variation of the electrical resistance of the buckypaper vs the deposited amount of isopropanol(drops).The variation of the resistance, for each peak, is the difference between the maximum and the starting value of the resistance.*

The growing up-steps in the resistance (Fig. 5 (b)) are expected considering that a wider buckypaper area is wetted. The error bars result from the instrumental incertitude and error propagation. We notice that the general up-trend of the resistance in Fig. 5 (a) is due to the fact that we did not wait long enough for the fully restore of the baseline after each drops' deposition. The large sensor response is connected to the extent of the interaction area between the carbon nanotubes and the alcohol. Consistently with our finding, a substantial increase of the resistance of entangled networks of pristine MWCNTs, exposed to vapors of linear alcohols (such as methanol, ethanol, 1-propanol, 1-butanol and 1-pentanol), was reported in reference [23].

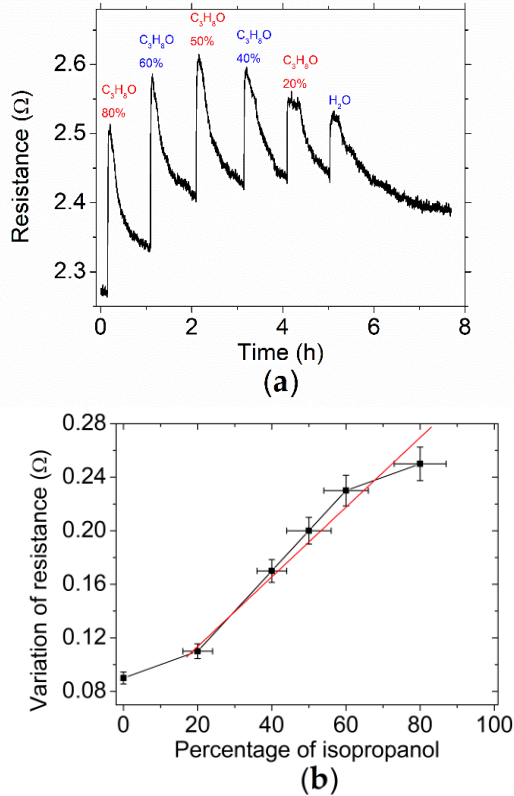

Fig. 6. *Sensor response to aqueous solutions with different isopropanol concentration. (a) Time-variations of the buckypaper resistance, induced by different isopropanol/water solutions.; (b) Change in the electrical resistance as a function of the concentration of isopropanol. The variation of the resistance is the difference between the maximum of the resistance and its starting value for each peak.*

When the liquid solution gets in contact with the CNT film, the solute molecules are polarized by the fringing electric fields radiating from the CNTs and are attracted to their surface, causing a change of the CNTs electric resistance. Molecules with higher dipole moment, such as water or alcohol, are attracted with higher efficiency [24]. Furthermore, polar COOH groups possibly created during the fabrication/purification process on the CNT surface could contribute to the adsorption efficiency of alcohol molecules by hydrogen bonding. Consequently, a charge transfer between the CNT to the adsorbed molecules, which reduces the intrinsic doping of the nanotubes, likely causes the observed increase of the electrical resistance of the buckypaper with the alcohol concentration. Furthermore, the molecules adsorbed on the carbon nanotube surface can induce a change in the structure of the buckypaper and make the CNT-CNT contacts looser, thereby increasing the tunneling barrier and the inter-tubes contact resistance. The above two mechanisms, that is the reduction of the MWCNT electron density and the increase of contact resistance between touching tubes, likely concur to cause the observed variation of resistance.

Next, we consider aqueous solutions consisting of different percentages of isopropanol and water (Figs. 6). Indeed, the sudden increase of resistance varies with the concentration of isopropanol in the solution. As shown in Fig. 6 (b), the sensor is capable of discriminating different isopropanol concentrations in the solution, exhibiting a monotonic variation of resistance with the percentage of isopropanol. We notice that this behavior is consistent with Fig. 5.

A natural problem that arises is the sensititvity of the buckypaper element to alcoholic beverages. Hence, we have checked the buckypaper resistance when the sample was exposed to alcoholic beverages, such as beer (5%), wine (12%), limoncello (24%), whisky (40%), xenta (75%), and pure alcohol (95%). The purpose was to figure out if the buckypaper was able to discriminate the concentration of alcohol in common beverages. To prevent the deposition of unwanted substances, present in the beverages, we cleaned the sample by water, which was poured on the buckypaper and let evaporate before each measurement. Fig. 7 shows the response to a single drop of several alcoholic beverages with augmenting alcohol volume content. Although the response might be contributed by chemicals present in the beverages, the linear increase of the resistance, consistently what has been observed for the isopropanol solutions, is mostly caused by the alcoholic concertation. The alcohol concentration response can be evaluated as $S = \frac{1}{\Delta R_0} \frac{d\Delta R}{dP} = 2.6$ % per unit of alcohol volume percentage (here P is the alcohol volume percentage, $\Delta R$ is the variation of resistance, and $\Delta R_0 = 58.7$ $m\Omega$ is the extrapolated variation of resistance for an alcohol-free beverage). This result indicates a high sensitivity of the MWCNT film to the alcoholic graduation of liquid solution and constitutes the proof-of-concept of the proposed sensor. Once optimized and properly calibrated, the sensor has great potentiality for real-life applications, such as for the control of the alcohol content in cosmetic products or in worldwide highly-consumed alcoholic beverages.

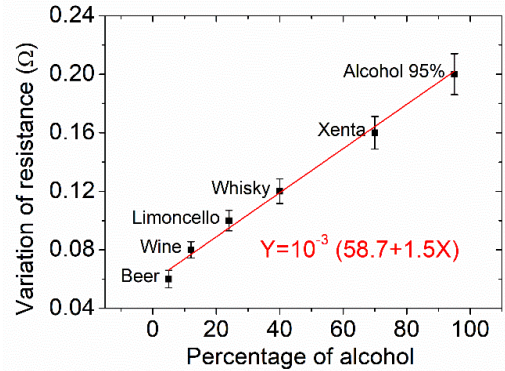

Fig. 7. *Electrical resistance measured when the sensor is exposed to a drop of several alcoholic beverages with increasing alcoholic content.*

**4. Conclusion:** In this work, we have demonstrated that a MWCNT film (buckypaper) responds to the alcoholic concentration of a liquid solutions. The observed increase of the electrical resistance with the alcohol concentration can be ascribed to stress due to either alcohol molecules captured around the carbon nanotubes surface or to charge transfer between the CNT and the alcohol molecules.

The buckypaper shows an ohmic behavior, with a positive temperature resistance coefficient, typical of a metal. We have proved that a steep rise of the resistance results from the exposure of the buckypaper to water drops with increasing content of isopropanol. The device shows repeatability and reusability, since its electrical resistance returns to the initial value when the solution drops evaporate. More importantly, we have demonstrated that the MWCNTs-based sensor has the potentiality to discriminate the concentration of alcohol in common alcoholic beverages. The sensor exhibits a linear and rapid response, and can be used to monitor a sudden change of the alcohol concentration in a solution.

This work paves the way towards practical application of multi-walled carbon nanotube films in the control of the alcoholic concentration of liquid substances.

**5. Acknowledgments:** We acknowledge the economic support of POR Campania FSE 2014–2020, Asse III Ob. Specifico l4, Avviso pubblico decreto dirigenziale n. 80 del 31/05/2016. We thank Prof. Ettore Varricchio for his support in this project.

**6. References:**

[1] S. Iijima: "Helical microtubules of graphitic carbon". Nature 1991, 354 (6348), 56–58, DOI: 10.1038/354056a0.

[2] F. Giubileo, L. Iemmo, G. Luongo, N. Martucciello, M. Raimondo, L. Guadagno, M. Passacantando, K. Lafdi, A. Di Bartolomeo: "Transport and field emission properties of buckypapersobtained from aligned carbon nanotubes". Journal of Materials Science 2017, 52 (11),6459-6468, DOI: 10.1007/s10853-017-0881-4.

[3] E. Dervishi, Z. Li, Y. Xu, V. Saini, A. R. Biris, D.Lupu, A. S. Biris: "Carbon Nanotubes: Synthesis, Properties, and Applications". Particulate Science and Technology 2009, 27 (2), 107-125, DOI: 10.1080/02726350902775962.

[4] M. Passacantando, F. Bussolotti, S. Santucci, A. Di Bartolomeo, F. Giubileo, L. Iemmo and A. M. Cucolo: "Field emission from a selected multiwall carbon nanotube". Nanotechnology 2008, 19, 395701, DOI:10.1088/0957-4484/19/39/395701.

[5] A. Di Bartolomeo, A. Scarfato, F. Giubileo, F. Bobba, M. Biasiucci, A. M. Cucolo, S. Santucci, M. Passacantando: "A local field emission study of partially aligned carbon-nanotubes by atomic force microscope probe". Carbon 2007, 45 (15), 2957-2971, DOI: 10.1016/j.carbon.2007.09.049.

[6] M. S. Dresselhaus, G. Dresselhaus, P. C. Eklund: "Science of Fullerenes and Carbon Nanotubes". Academic Press, 1996, New York, NY, USA, ISBN: 9780080540771.

[7] R. Saito, M. Fujita, G. Dresselhaus, M. S. Dresselhaus: "Electronic structure of chiral graphene tubules". Applied Physics Letters1992,60 (18), 2204–2206, DOI: 10.1063/1.107080.

[8] I. V. Zaporotskova, N. P. Boroznina, Y. N. Parkhomenko, L. V. Kozhitov: "Carbon nanotubes: Sensor properties. A review". Modern Electronic Materials 2016, 2 (4), 95–105, DOI: 10.1016/j.moem.2017.02.002.

[9] F. Giubileo, A. Di Bartolomeo, L. Iemmo, G. Luongo, F. Urban: "Field Emission from Carbon Nanostructures". Applied Sciences 2018, 8 (4), 526; DOI: 10.3390/app8040526

[10] N. Sinha, J. Ma, J. T. W. Yeow: "Carbon nanotube-based sensors". Journal of Nanoscience and Nanotechnology, 2006, 6(3), 573–590, DOI: 10.1166/jnn.2006.121.

[11] A. Calvi, A. Ferrari, L. Sbuelz, A. Goldoni, S.Modesti: "Recognizing Physisorption and Chemisorption in Carbon Nanotubes Gas Sensors by Double Exponential Fitting of the Response". Sensors, 2016, 16 (5), 731, DOI: 10.3390/s16050731.

[12] H. Chang, J. D. Lee, S .M. Lee, Y. H. Lee: "Adsorption of $NH_3$ and $NO_2$ molecules on carbon nanotubes". Applied Physics Letters 2001, 79, 3863–3865, DOI: 10.1063/1.1424069.

[13] M. Penza, G. Cassano, P. Aversa, F. Antolini, A. Cusano, A. Cutolo, M. Giordano, L. Nicolais: "Alcohol detection using carbon nanotubes acoustic and optical sensors". Appl. Phys. Lett. 2004, 85, 2379, pp. 3; DOI: 10.1063/1.1784872

[14] T. Someya, J. Small, Ph. Kim, C. Nuckolls, J. T. Yardley: "Alcohol Vapor Sensors Based on Single-Walled Carbon Nanotube Field Effect Transistors". Nano Letters 2003, 3 (7), pp 877–881 DOI: 10.1021/nl034061h

[15] S. Brahim, S. Colbern, R. Gump, A. Moser, L. Grigorian: "Carbon nanotube-based ethanol sensors". Nanotechnology 2009, 20 (23), 235502, pp. 7, DOI: doi:10.1088/0957-4484/20/23/235502

[16] C. Sutthinet, A. Sangnual, T. Phetchakul: "Alcohol sensor based on multi-wall carbon nanotube". Proceedings of the 2009 12th International Symposium on Integrated Circuits, Singapore 14-16 Dec 2009, IEEE 2009,11135977, ISBN: 978-9-8108-2468-6

[17] M. L. Y. Sin, G. C. T. Chow, G. M. K. Wong, W. J. Li, P. H. W. Leong, K. W. Wong: "Ultralow-power alcohol vapor sensors using chemically functionalized multiwalled carbon nanotubes". IEEE Transactions on Nanotechnology 2007, 6 (5), 571–577, DOI: 10.1109/TNANO.2007.900511.

[18] Y. Wang, J. T. W. Yeow: "A Review of Carbon Nanotubes-Based Gas Sensors". Journal of Sensor 2009, 493904, pp. 24, DOI:10.1155/2009/493904.

[19] A. Di Bartolomeo, M. Sarno, F. Giubileo, C. Altavilla, L. Iemmo, S. Piano, F. Bobba, M. Longobardi, A. Scarfato, D. Sannino, A. M. Cucolo, P. Ciambelli: "Multiwalled carbon nanotube films as small-sized temperature sensors". Journal of Applied Physics 2009, 105, 064518, pp. 6, DOI: 10.1063/1.3093680.

[20] F. Giubileo, A. Di Bartolomeo, M. Sarno, C. Altavilla, S. Santandrea, P. Ciambelli, A. M. Cucolo: "Field emission properties of as-grown multiwalled carbon nanotube films". Carbon 2011, 50, 163-169, DOI:10.1016/j.carbon.2011.08.015.

[21] H. Cui, X. Yan, M. Monasterio, F. Xing: "Effects of Various Surfactants on the Dispersion of MWCNTs–OH in Aqueous Solution". Nanomaterials 2017, 7(9), 262, DOI: 10.3390/nano7090262.

[22] M. Sarno, D. Sannino, C.Leone, P. Ciambelli: "Evaluating the effects of operating conditions on the quantity, quality and catalyzed growth mechanisms of CNTs". Journal of Molecular Catalysis A: Chemical 2012, 357, pp. 26- 38, DOI: 10.1016/j.molcata.2012.01.014

[23] R. Benlikaya, P. Slobodian, P. Riha, R. Olejnika: "The enhanced alcohol sensing response of multiwalled carbon nanotube networks induced by alkyl diamine treatment". Sensors and Actuators B2014, 201, 122–130, DOI: 10.1016/j.snb.2014.04.057 0925-4005.

[24] E. S. Snow, F. K. Perkins, E. J. Houser, S. C. Badescu, T. L. Reinecke: "Chemical Detection with a Single-Walled Carbon Nanotube Capacitor". Science 2005, 307, pp. 1942-1945, DOI: 10.1126/science.1109128

--------------------------------------------------------------------------------